\documentclass[doublecol]{epl2}
\usepackage{graphicx}
\usepackage{amsmath}
\usepackage{amssymb}
\usepackage{color}
\usepackage{bm}
\usepackage{subfigure}
\usepackage{epstopdf}
\newcommand{\beq}{\begin{equation}}
\newcommand{\eeq}{\end{equation}}
\newcommand{\beqa}{\begin{eqnarray}}
\newcommand{\eeqa}{\end{eqnarray}}

\title{Generic phases of cross-linked active gels: Relaxation, Oscillation and Contractility}

\author{S. Banerjee\inst{1} \and T.B. Liverpool\inst{2} \and M.C. Marchetti\inst{1,3}}

\institute{
  \inst{1} Physics Department, Syracuse University, Syracuse New York, 13244-1130, USA\\
  \inst{2} Department of Mathematics, University of Bristol, University Walk, Clifton, Bristol BS8 1TW, U.K.\\
  \inst{3} Syracuse Biomaterials Institute, Syracuse University, Syracuse New York, 13244-1130, USA}

\abstract{ We study analytically and numerically a generic continuum model of an isotropic active solid with internal stresses generated by non-equilibrium `active' mechano-chemical reactions.
Our analysis shows that  the gel can be tuned through  three  classes of  dynamical states by increasing motor activity:  a  constant  unstrained state of homogeneous density, a state where the local density exhibits sustained oscillations, and a steady-state which is spontaneously contracted, with a uniform mean density.
}

\begin{document}

\maketitle

\section{Introduction}

The mechanical properties of living cells are largely controlled by  a variety of filament-motor networks optimized for diverse physiological processes~\cite{AlbertsBook07}. In the presence of ATP, such networks are capable of generating controlled contractile forces and spontaneous oscillations.  These phenomena arise from the presence of groups of motor proteins, such as myosin II,  that convert the  chemical energy from ATP hydrolysis into mechanical work via a cyclic process of attachment and detachment to associated polar protein filaments, e.g. F-actin~\cite{HowardBook00}.
There is great variation in the  organization of actin, myosins, and other cross-linking proteins in the  actomyosin structures found in cells. Myofibrils  in striated muscle cells are examples of highly organized structures~\cite{AlbertsBook07}, composed of repeated subunits of actin and myosin, known as sarcomeres, arranged in series.  Each sarcomere consists  of actin filament of alternating polarity  bound at their pointed end by large clusters of myosins, known as myosin ``thick filaments".    The periodic structure of the myofibril  allows it to generate forces on large length scales  due to the collective dynamics of individual units of microscopic size, giving rise to muscle oscillation and contraction.  More difficult is to understand the origin of spontaneous oscillations and contractility  in cytoskeletal filament-motor assemblies that lack such a highly organized structure~\cite{AlbertsonYeh}.

Oscillations are for instance observed in many organisms during repositioning of the mitotic spindle from the cell center towards the cell pole when unequal cell division occurs~\cite{AlbertsonYeh}.
{\it In vitro} examples of such phenomena are  sustained cilia-like beatings in self-assembled bundles of  microtubules and dyneins~\cite{Sanchez2011} and spontaneous contractility  in isotropic reconstituted actomyosin networks with additional F-actin crosslinking proteins,  such as filamin or $\alpha$-actinin~\cite{Bendix2008,Koenderink2009}.

Theoretical models have shown that  spontaneous oscillations can arise from the collective action of groups of molecular motors coupled to a single elastic element~\cite{JulicherProst9597}. In these models the load dependence of the binding/unbinding kinetics of motor proteins breaks detailed balance and provides the crucial nonlinearity that  tunes the system through a Hopf bifurcation to a spontaneously oscillating state. This simple theoretical concept has been adapted to describe the beating of cilia and flagella~\cite{Camalet2000}, mitotic spindles during asymmetric cell division~\cite{GrillPlacais}, and spontaneous waves in muscle sarcomeres~\cite{GuntherKruse2007,GuntherKruse2010,Anazawa1992,Okamura1988}.

In a parallel development,   generic continuum  theories of active polar  `gels' have been constructed by suitable modification of the hydrodynamic equations of equilibrium liquid crystals to  incorporate the effect of activity. These continuum  models are capable of capturing some of the large-scale consequences of the internal contractile stresses induced by active myosin crosslinkers, including the presence of propagating actin waves in cells adhering to a substrate~\cite{Asano2009} and the retrograde flow  in the lamellipodium of crawling cells~\cite{JKPJ-PhysReports2007}. In these studies the acto-myosin network is modeled  as a Maxwell viscoelastic fluid,  with short-time elasticity and liquid response at {\em long times}~\cite{Kruse2005}.  The active viscoelastic liquid cannot, however, support sustained oscillations that require  low frequency, long wavelength  elastic restoring forces.

In this letter we consider a nonlinear version of  the generic continuum model of an isotropic active solid discussed recently by two of us~\cite{BanerjeeMarchetti2010}.
We go beyond the linear stability analysis  discussed in \cite{BanerjeeMarchetti2010} and show that  the presence of  nonlinearities leads  to both stable contracted and oscillatory states in different regions of parameters. The acto-myosin network is modeled as an elastic continuum, as appropriate for a cross-linked polymer gel embedded in a permeating viscous fluid, with elastic response at {\em long times} and liquid-like dissipation at short times.  The active solid model describes the various phases of acto-myosin systems as a function of motor activity, including  spontaneous contractility and oscillations. It provides a unified description of both phenomena and a minimal model relevant to many biological systems with motor-filament assemblies that behave as solids at low frequencies. When the active solid is isotropic, as assumed in the present work, the coupling to  a non-hydrodynamic mode provided by the binding and unbinding kinetics of motor proteins is essential to generate spontaneous sustained oscillations.

\begin{figure}
\begin{center}
          \includegraphics[width=0.26\textwidth]{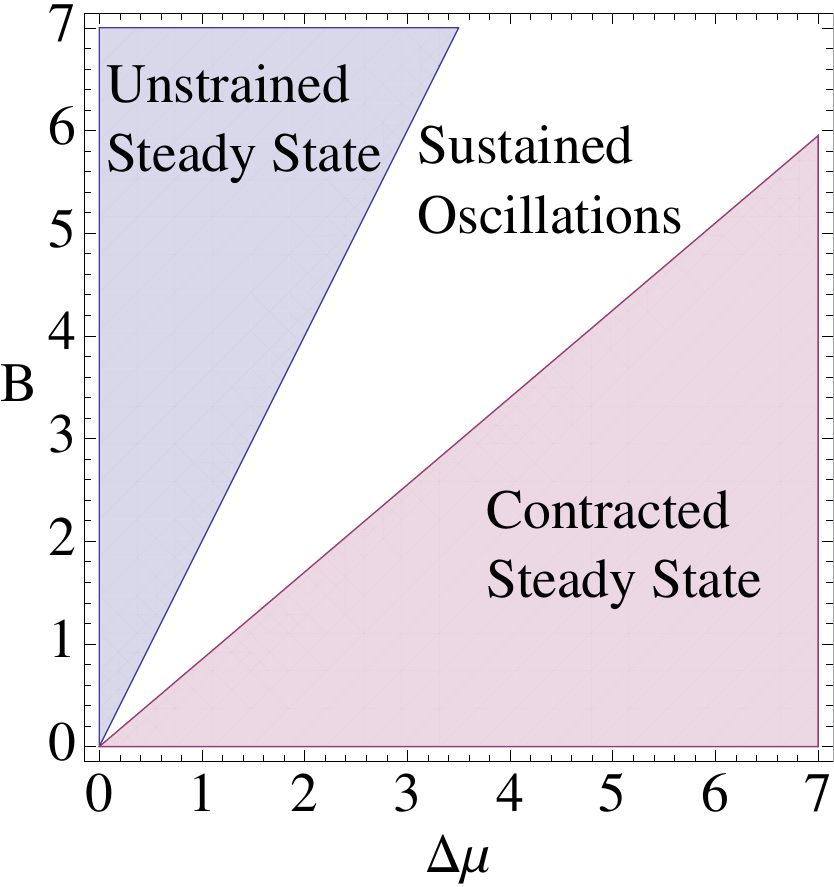}
\end{center}
\caption{(Color online)  ``Phases"  of the nonlinear active elastic gel obtained by varying the compressional modulus $B$ and the activity $\Delta\mu$.}
              \label{phase_system}
\end{figure}

 The main results are summarized in Fig.~\ref{phase_system}, where we sketch the steady states of the system as we tune the activity, defined as the difference  $\Delta\mu$ between the chemical potential of ATP and its hydrolysis products, and the compressional modulus $B$ of the passive gel.  For a fixed value of $B$ we find a regime where the active gel supports sustained oscillating states and a contracted steady state as $\Delta\mu$ is increased. For fixed $\Delta\mu$ spontaneous contractility is only observed below a critical network stiffness. This is consistent with experiments on  isotropic acto-myosin networks with additional cross-linking $\alpha$-actinin where  spontaneous contractility was seen only in an intermediate range of  $\alpha$-actinin concentration~\cite{Bendix2008}. Our model does not, however, yield a lower bound on $B$ below which the contracted state does not exist. This may be because, in contrast to the experiments where a minimum concentration of $\alpha$-actinin is required to provide integrity to the network, our system is always by definition an elastic solid, even at the lowest values of $B$.    The phase diagram resulting from our model also resembles the state space diagram of a muscle sarcomere  obtained  experimentally~\cite{Okamura1988}.

 Interestingly,
 we find that  a simplified dynamical system  obtained by a one-mode approximation to our  continuum theory corresponds to the half-sarcomere model proposed recently by G\"unther and Kruse~\cite{GuntherKruse2007}  for a particular set of parameters.  Our analysis shows, however,  that the phase behavior described above is generic  and can be expected in a wide variety of active elastic systems, as it relies solely on symmetry arguments. It provides a  unified description of both spontaneous  and oscillatory states and predicts their region of stability as a function of the elastic properties of the network and motor activity.

\section{Model}
In Ref.~\cite{BanerjeeMarchetti2010} two of us formulated, on the basis of symmetry arguments, a  phenomenological hydrodynamic model of a cross-linked gel (a network of actin filaments crosslinked by filamins or other "passive" linkers) under the influence of active forces exerted by clusters of crosslinking motor proteins (e.g., myosin II minifilaments).  Only linear terms were retained in the continuum equations in~\cite{BanerjeeMarchetti2010} where the linear modes of the system and their stability were analyzed. Here we consider a nonlinear continuum model and show how nonlinearities can stabilize oscillatory and contracted states.
The active gel consists of a polymer network  in a permeating fluid~\cite{BanerjeeMarchetti2010}. On length scales large compared to the network mesh size, the gel can be described as a continuum elastic medium, viscously coupled to a Newtonian fluid.   The model follows closely that formulated for a passive gel~\cite{Levine2001}. We focus on compressional deformations and consider for simplicity a one-dimensional model. We assume the permeating fluid to be incompressible and consider the case of small volume fraction of the gel. In this limit the permeating fluid simply provides a frictional drag to the polymer network.  The hydrodynamic description is then obtained in terms of two conserved fields: the density $\rho(x,t)$ of the gel and the one-dimensional displacement field, $u(x,t)$. The momentum density of the gel is not conserved due to drag exerted by the permeating fluid. In addition, density variations are determined by the local strain, with
$\delta\rho=\rho-\rho_0=-\rho_0\partial_xu$ and $\rho_0$ the equilibrium mean density.
The elastic free energy density of the gel can be expanded in the strain $s=\partial_x u$ about the state $s=0$. To  describe the possibility of swelling and collapse of a  gel (even a passive one) one needs to keep terms up to fourth order in $s$  in the elastic free energy density (dropping  a linear term that can be eliminated by redefining the ground state)~\cite{TanakaLubensky}
 \beq
 \label{fe}
 f_e=\frac{B}{2}s^2 + \frac{\alpha}{3} s^3 + \frac{\beta}{4}s^4\;,
 \eeq
where $B$ is the longitudinal compressional modulus of the gel and $\alpha,\beta>0$ are phenomenological parameters capturing the effects of many-body interactions and nonlinear elasticity of the components~\cite{StormGardel}. 

It is straightforward to show that within mean-field theory the free energy given in Eq.~\eqref{fe} yields  a line of first order phase transitions at $B\equiv B_c=2\alpha^2/(9\beta)$ between an unstrained state with $s=0$ for $B>B_c$ and a strained state with finite $s$ for $B<B_c$. The stable strained state is one of higher density ($s<0$) for $\alpha>0$, corresponding to a contracted state, and one of lower density ($s>0$) for $\alpha<0$, corresponding to an expanded state. The transition line terminates at a critical point at $\alpha=0$.

Activity is induced by the presence of a concentration $c(x,t)$ of bound active cross-linkers that undergo a cyclic binding/unbinding transformation fueled by ATP, exerting forces on the gel.
The  dynamics of the active gel on time scales larger than the Kelvin-Voigt viscoelastic relaxation time is described by  coupled equations for $u(x,t)$ and $c(x,t)$, given by
\begin{eqnarray}
&&\Gamma\partial_t u=\partial_x\sigma_e -\partial_x p_a \label{u-eq}\;,\\
&&\partial_tc=- \partial_{x}(c\partial_t{u}) - k(s)(c -c_0)\label{motor1}
\end{eqnarray}
where $\sigma_e=\frac{\partial f_e}{\partial s}=Bs+\alpha s^2+\beta s^3$ is the elastic stress, $\Gamma$ is a friction constant describing the coupling of the polymer network to the permeating fluid, and $p_a(\rho,c)$ is the active contribution to the pressure, describing the isotropic part of the active stresses induced by myosins. Equation \eqref{motor1} allows for convection of bound motors by the gel at speed $\partial_tu$ and incorporates the binding/unbinding dynamics, with  $k(s)$ the strain-dependent motor unbinding rate and
$c_0$ the equilibrium concentration of bound motors.
 Since highly non-processive motor proteins such as myosins are on average largely unbound, we neglect the dynamics of free motors that  provide an infinite motor reservoir.

The active pressure is taken to be linear in the rate $\Delta\mu $ of ATP consumption, with $p_a=\Delta\mu~\zeta(\rho,c)$.  This is a reasonable approximation for weakly active systems,  where the number of active elements make up a small fraction of the total mass of the gel, as is the case in most experiments.~\footnote{We note that the effects of nonlinearities in $\Delta \mu$  can taken account of by  expanding in $\delta \mu = \Delta \mu -  \Delta\mu_0$ about a stationary state with finite $\Delta\mu_0$.} We then expand 
\beq
\begin{split}
\label{pressure}
\zeta(\rho,c) \simeq &\zeta_0- \zeta_1\delta\tilde\rho-  \zeta_2 \phi - \zeta_3 \left(\delta\tilde\rho\right)^2 +
\zeta_4\delta\tilde\rho\phi \\ & +\zeta_5 \phi^2 + \zeta_6 \left(\delta\tilde\rho\right)^3...
\end{split}
\eeq
with $\delta\tilde\rho=\delta\rho/\rho_0$ and $\phi=(c-c_0)/c_0$ the fluctuations in the gel and bound motor concentrations, respectively, and
 all parameters $\zeta_i$ defined  positive. The positive sign of  $\zeta_1$ and $\zeta_2$  corresponds to a contractile acto-myosin system and describes the reduction in the longitudinal stiffness of the gel  from contractile forces exerted by motor proteins. The parameters $\zeta_3$, $\zeta_4$ and $\zeta_5$  describe excluded volume effects. A positive $\zeta_3$  favors contracted over expanded states. A positive $\zeta_6$  guarantees the stability of the network in regions of negative effective  compressional modulus.

 There are several sources of nonlinearities in  Eqs. (\ref{u-eq},\ref{motor1}):  the nonlinear strain dependence of the elastic free energy, the nonlinear terms in the active pressure, and the dependence of the motor   unbinding rate $k$ on the load force on bound motors, which in turn is proportional to the strain $s$ of the elastic gel. We  assume an exponential dependence of the form $k(s)=k_0e^{\epsilon s}$~\cite{Parmeggiani2001}, with $k_0$ the unbinding rate of unloaded motors and $\epsilon$ a dimensionless parameter  determined by   microscopic properties of the  motor-filament interaction. In the following we expand the unbinding rate for small strain as
\begin{math}
\label{ku}
k(s) \simeq k_{0}\left[1 + \epsilon s + \frac{\epsilon^2}{2} s^2+{\cal O}(s^3)\right].
\end{math}
Keeping higher order terms in $\epsilon$ does not change qualitatively the behavior described below. Finally, the first term on the rhs of Eq. (\ref{motor1}) contains a convective nonlinearity that  does not affect the key features of dynamics observed and  will be neglected in most of the following.

The equations for the active gel can then be written as
%
\begin{subequations}
\begin{gather}
\Gamma\partial_t u=\partial_x\sigma_e^a  + \Delta\mu \partial_x \left[\zeta_2\phi + \zeta_4s\phi - \zeta_5 \phi^2\right]    \;,\label{u2}\\
\partial_t\phi=- \partial_{x}\left[\left(1+\phi\right) \partial_t{u}\right] - k_{0}\left[1 + \epsilon s+ \frac{\epsilon^2}{2} s^2 \right] \phi \;,\label{motor2}
\end{gather}
\end{subequations}
%
where $\sigma_e^a=B_as+\alpha_a s^2+\beta_a s^3$, with renormalized elastic constants,  $B_a=B-\zeta_1\Delta\mu$, $\alpha_a=\alpha + \zeta_3\Delta\mu$ and $\beta_a=\beta+\zeta_6\Delta\mu$.

We render our equations dimensionless by letting $u\rightarrow u/L$ and $t\rightarrow tk_0^{-1}$. We then  define dimensionless parameters as $\tilde{B}=\frac{B}{\Gamma k_0 L^2}$, $\tilde{\alpha}=\frac{\alpha}{\Gamma k_0 L^2}$, $\tilde{\beta}=\frac{\beta}{\Gamma k_0 L^2}$, $\Delta\tilde{\mu}=\frac{\Delta\mu}{\Gamma k_0 L^3}$ and $\tilde\zeta_i=\zeta_i L$. In the following we drop the tilde to simplify notation and all quantities should be understood as dimensionless.

\section{Linear stability analysis}

There are three steady state solutions of Eqs.~ (\ref{u2},\ref{motor2}): an unstrained state, with $(u_s, \phi_s)=(0,0)$, and two strained states, with $(u_s, \phi_s)=(s_\pm x ,0)$ and $s_\pm=\left(-\alpha_a\pm\sqrt{\alpha_a^2-4B_a\beta_a}\right)/2\beta_a$, provided $B_a < \alpha_a^2/4\beta_a$. For concreteness we choose $\alpha_a,\beta_a>0$. Then if $B_a>0$, $s_\pm<0$ and $\delta\tilde\rho_\pm>0$, so that both strained solutions correspond to contracted states.
If $B_a<0$, then $s_-<0$ and $s_+>0$,  corresponding to contracted and expanded states, respectively.

In this section we examine the linear stability of each of these states. Letting $s=s_0+\delta s$, where $s_0=(0,s_\pm)$ represents the constant value of strain in the steady state and $\delta s=\partial_x\delta u$, the linearized equations for the displacement and motor concentration fluctuations are given by
\begin{subequations}
\begin{gather}
\partial_t \delta u={\cal B}\partial_x^2\delta u+{Z}\partial_{x}\phi  \;,\label{ul1}\\
\partial_t\phi=- \partial_x \partial_t \delta u -\kappa\phi \;,\label{motorl1}
\end{gather}
\end{subequations}
where
\begin{subequations}
\begin{gather}
{\cal B}=B_a+2\alpha_as_0+3\beta_as_0^2\;,\\
{Z}=\Delta\mu(\zeta_2+\zeta_4s_0)\;,\\
\kappa=1+\epsilon s_0+\epsilon s_0^2/2\;.
\end{gather}
\end{subequations}
Looking for solution of the form $\delta u,\phi\sim e^{zt+iqx}$ we find that the dynamics of fluctuations is controlled by two eigenvalues, with dispersion relations
\begin{eqnarray}
\label{modes1}
z_{u,\phi}(q)=-\frac{b(q)}{2}
\pm\frac12
\sqrt{\left[b(q)\right]^2-4\kappa{\cal B}q^2}
\end{eqnarray}
where $b(q) = \kappa + ({\cal B}-Z)q^2$ and $z_u(q)$ and $z_\phi(q)$ correspond to the root with the plus and minus signs, respectively.

If motor fluctuations are neglected ($\phi=0$), corresponding to letting $Z=0$ in Eq.~\eqref{modes1}, one finds $z_u=-{\cal B}q^2$ and the linear stability of the steady states is entirely determined by the sign of ${\cal B}$, with ${\cal B}>0$, corresponding to a linearly stable state.  When $\phi=0$ the problem is equivalent to an equilibrium problem,  with $q^2{\cal B}$  the curvature  of a free energy of the form given in Eq.~\eqref{fe}, albeit with elastic constants renormalized by activity. In this case the unstrained state $s=0$ is the global minimum of the free energy for  $B_a>2\alpha_a^2/(9\beta_a)$, while the contracted state $s=s_-$ is the global minimum for $B_a<2\alpha_a^2/(9\beta_a)$. The line $B_a=2\alpha_a^2/(9\beta_a)$ defines a line of first order phase transitions in the $(B,\Delta\mu)$ (see Fig.~\ref{phase_nomotors}, left frame). If, on the other hand, we consider the problem dynamically, the condition of stability of linear fluctuations given by ${\cal B}>0$ is a necessary, but not sufficient one to identify the stable steady states of the system, as multiple fixed points with different basins of attraction coexist in the same region of parameters. In particular,  both the unstrained $s=0$ state and the contracted $s=s_-$ state are linearly stable for $B_a< B<\alpha_a^2/(4\beta_a)$, while both the contracted and expanded states ($s=s_\mp$) are stable for $B_a<0$. If we assume that among these linearly stable states the system selects dynamically the steady state with the fastest decay rate $q^2{\cal B}$, then we recover the linear phase diagram of Fig.~\ref{phase_nomotors} obtained form the equilibrium analysis.
\begin{figure}
\begin{center}
          \label{phase_nomotors}
\includegraphics[width=0.24\textwidth]{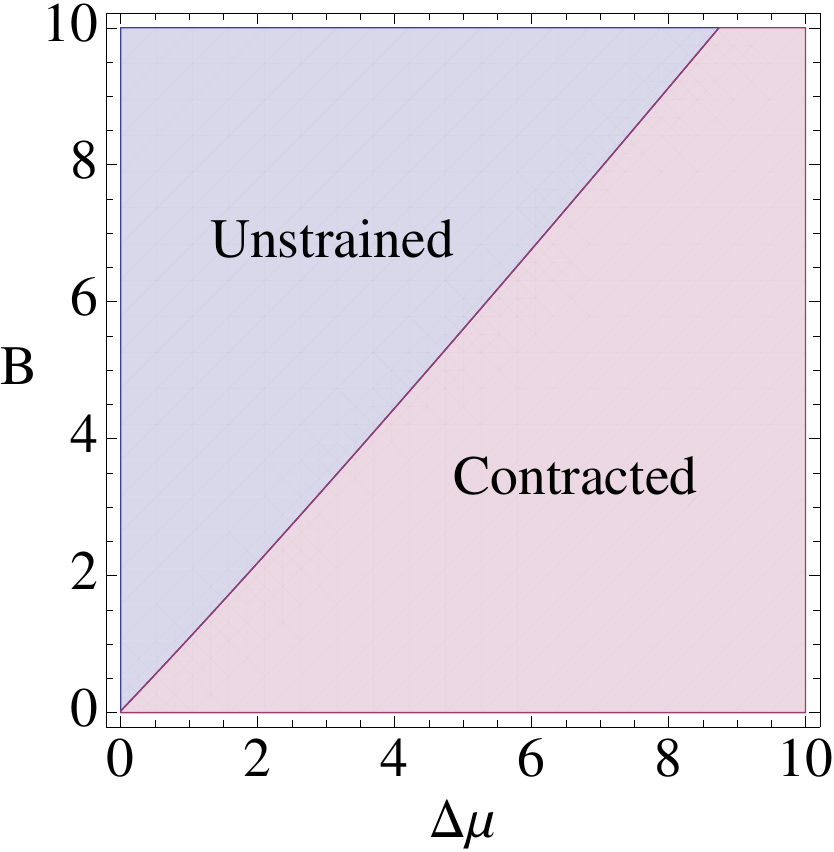}\includegraphics[width=0.24\textwidth]{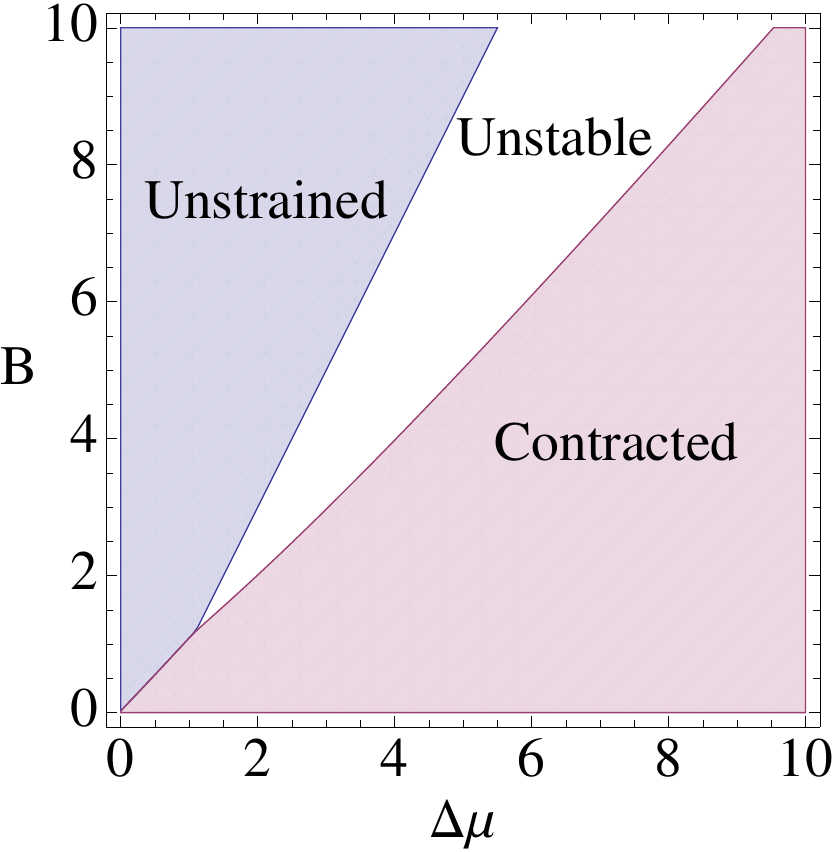}
 \end{center}
 \caption{(Color online) Phase diagram obtained by analyzing the linear stability of the modes. The left frame is  for $\zeta_2=\zeta_4=0$, corresponding to  $\phi=0$. The right frame is obtained including motor fluctuations, with  $\zeta_2=\zeta$ and $\zeta_4=\eta \zeta$. The other parameter values are $\alpha=0.1$, $\beta=0.1$, $\zeta_1=\zeta=1$, $\zeta_3=\eta \zeta$, and $\zeta_6=\eta^2\zeta$,  with $\eta=0.1$.}
 \label{phase_nomotors}
 \end{figure}

When the coupling to motor concentration fluctuations is included,  we find a region of parameters where $Re\left[z_{u,\phi}\right]>0$ for all three homogeneous solutions $s=(0,s_\pm)$. This is the white region in Fig.~\ref{phase_nomotors} (right frame), where the dynamical system is linearly unstable. The instability occurs in a region where the modes are complex, describing oscillatory states, and $Z-\kappa/q^2>{\cal B}>0$.  The linear stability diagram for the system is shown in Fig.~\ref{phase_nomotors} for the shortest wavevectors $\sim 1/L$, with $L$ the system size. The phase diagram is constructed again by assuming that the system will relax to linearly stable states characterized by the fastest relaxation rates. To linear order, the coupling to motor fluctuations yields oscillatory or propagating (as opposed to purely diffusive) fluctuations. Some of these oscillatory states are linearly unstable in a region of parameters, as shown in the phase diagram in Fig.~\ref{phase_nomotors}.

Next we consider the effect  of nonlinearities by first examining  a minimal model that only retains the longest wavelength mode in the Fourier expansion of the nonlinear equations, Eqs. (\ref{u-eq},\ref{motor1}),  and then comparing the latter to the numerical solution of the full nonlinear partial differential equations.

\section{One-mode model}
We begin by  incorporating only the  nonlinearities in the unbinding rate, while neglecting convective, elastic and pressure nonlinearities.
We impose boundary conditions $[\partial_x u]_{x=0}=[\partial_x u]_{x=L}=0$ and $[\phi]_{x=0}=[\phi]_{x=L}=0$, and seek a solution of Eqs.~(\ref{u-eq},\ref{motor1}) in the form of a Fourier series as, $u(x,t)=\sum_{m=0}^{\infty} u_m(t) \cos{(\hat{m} x)}$ and $\phi(x,t)=\sum_{m=1}^{\infty} \phi_m(t) \sin{(\hat{m}x)}$,
 where, $\hat{m}=m\pi/L$. We perform a 1-mode Galerkin truncation, and only consider the dynamics of the first nontrivial mode, $m=1$ (setting $u_m=\phi_m=0, \; \forall \; m \ne 1$). This corresponds to  approximating the system by only its  longest wavelength excitations, ignoring all the shorter wavelength modes.

 The coupled equations for $u_1$ and $\phi_1$ are given by
\begin{subequations}
\begin{gather}
\dot{u}_1=-B_a \pi^2u_1 +\zeta_2\Delta\mu\pi\phi_1\;,\label{u-one}\\
\dot{\phi}_1=\pi \dot{u}_1 -\left(1 - \frac{8}{3}\epsilon u_1 + \frac{3}{8}\pi^2\epsilon^2 u_1^2\right)\phi_1  \;.\label{motor-one}
\end{gather}
\end{subequations}
These equations can be recast into an effective second order differential equation for $u_1$ that has the structure of the equation for a Van der Pol oscillator~\cite{Strogatz1994} coupled to a nonlinear spring, of the form  $\ddot{u}_1  +\dot{u}_1\left[\lambda - f( u_1) \right] +  u_1B_a\left[1-f(u_1)\right]=0$, with $f(u_1)=\frac{8\epsilon}{3} u_1 -\frac{3\epsilon^2\pi^2}{8} u_1^2$ and  friction $\lambda=\pi^2(B_a-\zeta_2\Delta\mu)+1$.
Equations \eqref{u-one} and \eqref{motor-one}  admit one fixed point $(u_1,\phi)=(0,0)$. Linear stability analysis about this null fixed point  shows that the fixed point
is unstable (a repeller) when $\lambda < 0$ and is stable for positive $\lambda$.  From a global analysis~\cite{Strogatz1994} of Eqs. \eqref{u-one} and \eqref{motor-one} with the  $\epsilon$ nonlinearity,  we find that the existence of the unstable fixed point signals the appearance a stable limit cycle as $\lambda$ crosses zero. In other words the system undergoes a supercritical Hopf bifurcation at $\Delta\mu=\Delta\mu_{c1}=(B+\pi^{-2})/(\zeta_1+\zeta_2)$, with sustained oscillations for $\Delta\mu>\Delta\mu_{c1}$ and stable spirals or nodes~\cite{Strogatz1994} in  the two-dimensional, $(u_1,\phi_1)$ phase space,  otherwise.

We now consider the role of pressure and elastic nonlinearities, while letting $\epsilon=0$. Within the one-mode model
\begin{figure}
\begin{center}
\includegraphics[width=0.25\textwidth]{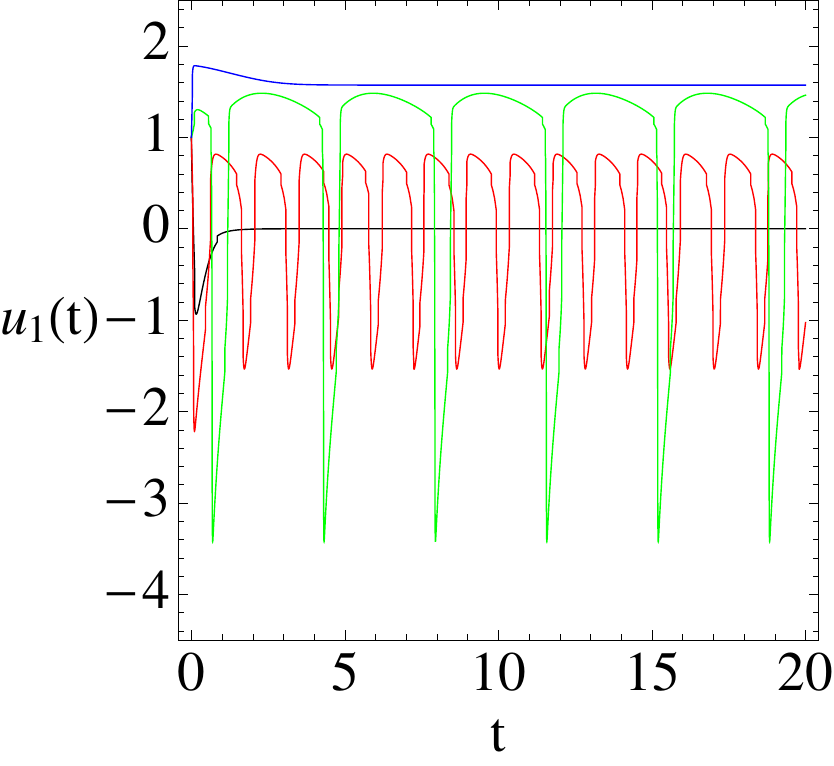}
\includegraphics[width=0.23\textwidth]{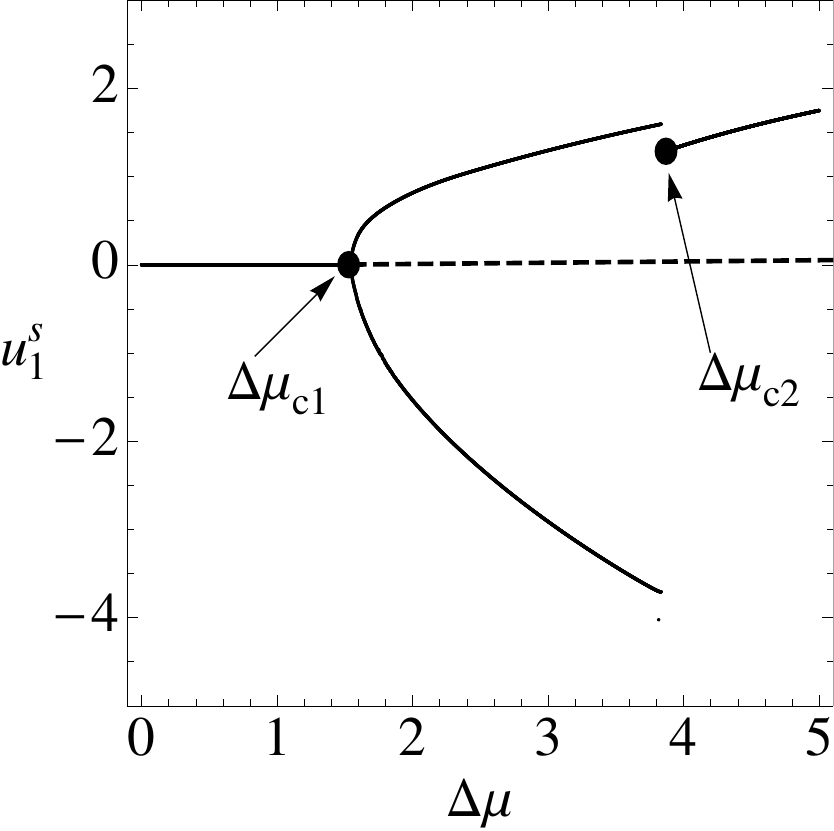}
\end{center}
\caption{(color online) Left frame : Time evolution of $u_1(t)$ for various values of activity : $\Delta\mu=1.0$ (black), corresponding to an unstrained steady state, $\Delta\mu=2.0$ (red) and $\Delta\mu=3.5$ (green), corresponding to the region of sustained oscillations and $\Delta\mu=4.5$ (blue), corresponding to the contracted steady state. Right frame: bifurcation diagram obtained from the 1-mode model. The figure shows a plot of the value $u_1^s$ of $u_1$ at long times ($t=100$)  vs $\Delta\mu$. A supercritical Hopf bifurcation occurs at $\Delta\mu=\Delta\mu_{c1}\simeq1.5$. As one increases $\Delta\mu$ within the range $\Delta\mu_{c1}<\Delta\mu<\Delta\mu_{c2}$, the amplitude of oscillations grows continuously. At $\Delta\mu=\Delta\mu_{c2}$ the limit cycle disappears and the system settles into a contracted steady state for $\Delta\mu>\Delta\mu_{c2}$. The dashed line indicates an unstable fixed point.
Parameters: $B=3.0$, $\epsilon=0.1$. Other parameter values are same as in Fig.~\ref{phase_nomotors}.}
\label{bifur_1mode}
\end{figure}
we find that in this case if  $\beta_a=\alpha_a=0$ and $B_a>0$ there is again only one fixed point at $(u_1,\phi)=(0,0)$ and the linear stability analysis is identical to that of the model with only rate nonlinearity. In other words, if $\epsilon=0$, the pressure nonlinearities do not stabilize the system for $\lambda<0$. When $\alpha_a\not=0,\beta_a\not=0$ and $B_a<16\alpha_a^2/27\beta_a\pi^2$, we have two new nonzero fixed points  $(u_1,\phi)\equiv (u_\pm,0)$, describing a contracted and an expanded state of the gel, respectively, with
\begin{equation}
\label{u-pm-onemode}
u_\pm=\frac{8\alpha_a}{9\pi^2\beta_a}\pm\frac{2}{3\beta_a\pi^2}\sqrt{\frac{16\alpha_a^2}{9}-3\pi^2\beta_a B_a}\;.
\end{equation}
Since $\delta\rho\simeq\pi\rho_0 u_1$, $u_+$ corresponds to a contracted state whereas $u_-$ corresponds to an expanded state when $B>\zeta\Delta\mu$. The contracted/expanded state is linearly unstable when
\begin{math}
\label{lambda-star}
\lambda^*_\pm=-B_a+\frac{8\alpha_a u_\pm}{3}-\frac{9\beta_a\pi^2u_\pm^2}{4}+\zeta_2-\frac{8\zeta_4u_\pm}{3}-\frac{1}{\pi^2}>0\;.
\end{math}
For $\lambda^*_\pm<0$ the nonzero fixed points are unstable if $B_a>16\alpha_a^2/27\beta_a\pi^2$. When $B_a<16\alpha_a^2/27\beta_a\pi^2$ and $\lambda^*_\pm<0$ the nonzero fixed points are stable and the orbits in the two-dimensional, $(u_1,\phi_1)$ phase-space
describe nodes or spirals settling at long times to $(u_\pm, 0)$~\cite{Strogatz1994}. When $\lambda^*_\pm>0$, the contracted/expanded states are linearly unstable. The nonlinearities in the active pressure stabilize these unstable states into stable asymmetric limit cycles. A supercritical Hopf bifurcation occurs when $\Delta\mu>\Delta\mu_{c1}=(B+\pi^{-2})/(\zeta_1+\zeta_2)$, which terminates to a stable contracted steady state for $\Delta\mu>\Delta\mu_{c2}$ ($\Delta\mu_{c2}$ is determined by the solution of $\lambda^*_\pm(B,\Delta\mu)=0$).

It is instructive to note that when $\alpha_a=\beta_a=0$, our 1-mode model corresponds to that of a half-sarcomere derived in Ref.~\cite{GuntherKruse2007}.  The bifurcation diagram for the complete 1-mode model is shown in Fig.~\ref{bifur_1mode}, with all the nonlinearities incorporated. The diagram summarizes the steady state crossovers of the system as we increase $\Delta\mu$ keeping $B$ fixed. Finally, the steady states of the nonlinear gel within the 1-mode approximation are shown in the phase diagram in Fig.~\ref{phase_pde} (indicated by dashed lines), in the $(B,\Delta\mu)$ plane. The unstrained state of the gel is stable when $\lambda>0$ whereas the contracted state is stable for $\lambda^*_+>0$. When $\lambda<0$ and $\lambda_+^*<0$ the gel exhibits sustained oscillations.

\section{Numerical analysis of the Continuum Nonlinear Model}
In this section we discuss the numerical solution of the nonlinear PDEs given in Eqs.~(\ref{u2},\ref{motor2}) (but with no convective nonlinearities)
with boundary conditions $\left[\partial_x u(t)\right]_{x=0,L}=0$ and $\left[\phi\right]_{x=0,L}=0$ and an initial strained state. We spatially discretize the PDEs using finite difference method and then integrate the resulting coupled ODEs  using the rkf45 method.

\begin{figure}
\begin{center}
\includegraphics[width=0.35\textwidth]{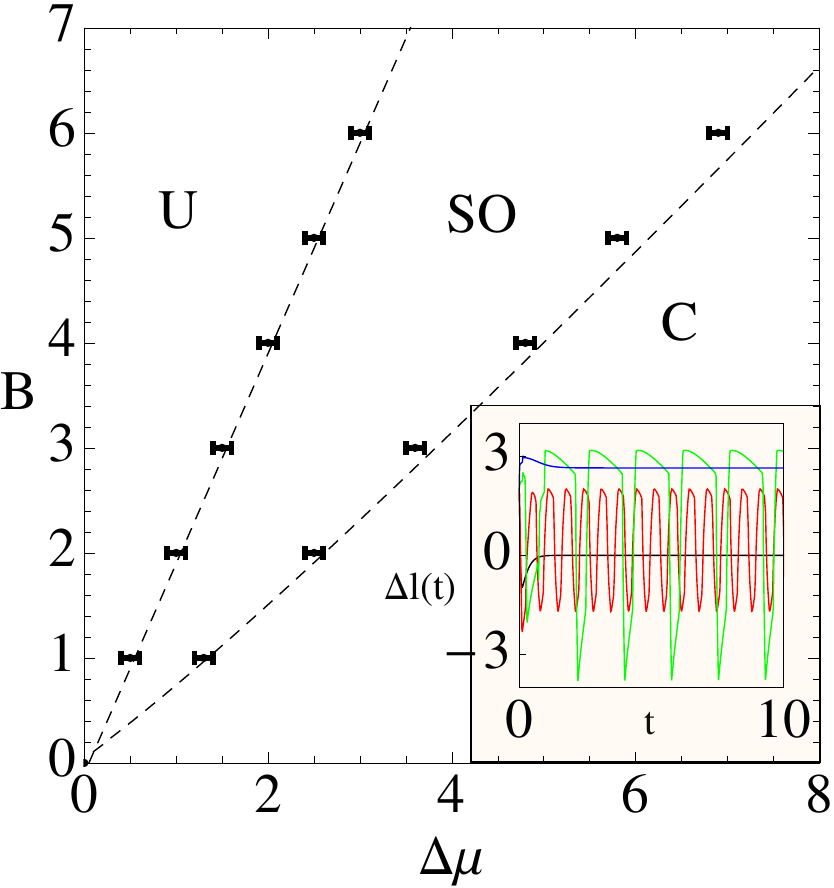}
\end{center}
\caption{(Color online) Phase diagram in the $(B,\Delta\mu)$ plane for the continuum nonlinear model described by Eqs. \eqref{u2} and \eqref{motor2}. U : Unstrained, SO : Sustained Oscillations, C : Contracted.   The points are obtained by numerical solution of the full PDE's; the error bars are determined by the step size used in the $\Delta\mu$ increments.  The dashed lines indicate the 1-mode phase boundaries. Inset : Plot of extension $\Delta\ell(t)=u(0,t)-u(L,t)$ for $B=3.0$ and $\Delta\mu=1.0$ (black), $\Delta\mu=2.0$ (red), $\Delta\mu=3.5$ (green) and $\Delta\mu=4.5$ (blue). Parameter values are same as in Fig.~\ref{bifur_1mode}.}
\label{phase_pde}
\end{figure}

The behavior of the system as we vary $B$ and $\Delta\mu$, keeping all other parameters constants, is summarized in the numerically constructed phase diagram, shown in Fig.~\ref{phase_pde}. Assuming $\zeta_1=\zeta_2=\zeta$, the system settles into an unstrained state for $B>B_{c1}\simeq2\zeta\Delta\mu$ and Hopf bifurcates to sustained oscillations for $B<B_{c1}$. We note that the phase boundary in the 1-mode model $B=2\zeta\Delta\mu-1/\pi^2$, although different from the numerical phase boundary $B=B_{c1}$, lies within the error bar of the numerics. For $B<B_{c2}\simeq 0.85 \zeta\Delta\mu$ the system settles into a stable contracted state. The basin of attraction of the expanded state is much smaller than that of the contracted state. This is due to the choice of the sign of the  coupling constants, $\alpha_a$ and $\zeta_4$. The  regimes predicted by our continuum phenomenological model may be used  to classify the behaviour seen in a number of  experimental systems~\cite{Okamura1988,Bendix2008,Koenderink2009}. To estimate the parameter  in real systems we assume $\Gamma\sim\eta/\xi^2$, with $\eta$ the viscosity of the permeating fluid and $\xi$ the gel mesh size~\cite{Levine2001} and use experimental values of $B$. For muscle fibers, using   $B\sim 2$ kPa~\cite{Magid1985},  $\eta \sim 2\times 10^{-3}Pa~s$~\cite{HowardBook00}, $\xi=0.5~\mu m$, $L\sim 100\ \mu m$ and $k_0^{-1}=40\ ms$, we estimate the dimensionless modulus $\tilde{B}=B/\Gamma L^2 k_0$ as $\tilde{B} \sim 1$. For isotropic cross-linked actomyosin gels only direct measurements of the low frequency shear modulus $G$ are available, with $G\sim 1-10Pa$~\cite{Koenderink2009}. It has been argued, however, that these networks may support much higher compressional forces, of the order of buckling forces on the scale of the mesh-size, yielding  a value of $B$ comparable to that of muscle fibers~\cite{Chaudhuri2006}. For $B\sim 1-10^3 Pa$ and $\eta$ comparable to that of water, we obtain $\tilde{B}\sim 10^{-3}-1$. The active coupling $\zeta\Delta\mu$ can be estimated as~\cite{GuntherKruse2007}  $\zeta\Delta\mu \sim \xi n_b k_m\Delta_m$, where $n_b$ is the number density of bound motors, $k_m$ is the stiffness of the myosin filaments and $\Delta_m$ is the steady state stretch of the bound motor tails. Using $n_b \sim 10^2-10^4~\mu m^{-3}$,  $k_m=4 pN/nm$ and $\Delta_m\sim 1\ nm$, we obtain $\zeta\Delta\mu \sim 0.1-10$ in dimensionless units. In the phase diagram in Fig.~\ref{phase_pde}, these parameter values would put muscle fibers in the regions U, SO and C respectively as we increase $n_b$. Isotropic actomyosin networks may have a much lower value of $\tilde{B}$, possibly corresponding to the contracted region. Indeed, while spontaneous contractility has been observed in vitro in reconstituted actomyosin networks~\cite{Bendix2008}, no experimental evidence has yet been put forward of spontaneous oscillations in these systems.

\section{Conclusion}
We have presented  a generic continuum model of a cross-linked active gel which can be used to describe a wide variety of isotropic elastic active systems.  We find an elastic active gel system can,  in general, be tuned through  three classes of dynamical states by increasing motor activity:  an unstrained steady state of homogeneous constant density, a state where the local density exhibits sustained oscillations, and a spontaneously contracted steady state, with a uniform mean density. The continuum model is not strictly hydrodynamic due to presence of the fast variable $\phi$, describing the  dynamics of motor proteins. The motor binding/unbinding kinetics plays a crucial role in generating oscillating states at finite wavevectors. The one-mode  model is in excellent qualitative agreement with the results described by the solutions to the nonlinear continuum equations and is comparable to a variety of one dimensional models for active elastic systems. Quantitative agreement in the phase boundaries between the one-mode model and the continuum model fails due to the non hydrodynamic nature of the model and we expect the oscillatory and/or contractile behaviour to depend on system size~\cite{Thoresen}.

\acknowledgments
MCM and SB were supported by the National Science Foundation through awards DMR-0806511 and DMR-1004789. TBL acknowledges the support of the EPSRC under grant EP/G026440/1.


\begin{thebibliography}{10}
\expandafter\ifx\csname url\endcsname\relax\def\url#1{\texttt{#1}}\fi

\bibitem{AlbertsBook07}
\Name{Alberts B. {\it et al.}}
  \Book{Molecular biology of the cell} (Garland, New York) 2007.

\bibitem{HowardBook00}
\Name{Howard J.} \Book{Mechanics of motor proteins and the cytoskeleton}
  (Sinauer Associates Sunderland, MA) 2001.

\bibitem{AlbertsonYeh}
\Name{Albertson D.} \REVIEW{Dev. Biol.}{101}{1984}{61};
\Name{Yeh E. \it{et al.}}
  \REVIEW{Mol. Bio. Cell }{11}{2000}{3949}.

\bibitem{Sanchez2011}
\Name{Sanchez T., Welch D., Nicastro D. \and Dogic Z.} \REVIEW{Science}{333}{2011}{456}.

\bibitem{Bendix2008}
\Name{Bendix P.~M. {\it et al.}} \REVIEW{Biophys.
  J.}{94}{2008}{3126}.

\bibitem{Koenderink2009}
\Name{Koenderink G.~H. {\it et al.}} \REVIEW{Proc. Nat. Acad. Sci.}{106}{2009}{15192}.

\bibitem{JulicherProst9597}
\Name{J{\"u}licher F. \and Prost J.} \REVIEW{Phys. Rev. Lett.}{75}{1995}{2618};
\Name{J{\"u}licher F. \and Prost J.} \REVIEW{Phys. Rev. Lett.}{78}{1997}{4510}.

\bibitem{Camalet2000}
\Name{Camalet S. \and J{\"u}licher F.} \REVIEW{New J. Phys.}{2}{2000}{24}.

\bibitem{GrillPlacais}
\Name{Grill S.~W., Kruse K. \and J{\"u}licher F.} \REVIEW{Phys. Rev. Lett.}{94}{2005}{108104};
\Name{Pla\ifmmode~\mbox{\c{c}}\else \c{c}\fi{}ais P.-Y., Balland M., Gu\'erin T., Joanny J.-F. \and Martin P.} \REVIEW{Phys. Rev. Lett.}{103}{2009}{158102}.

\bibitem{GuntherKruse2007}
\Name{G{\"u}nther S. \and Kruse K.} \REVIEW{New J. Phys.}{9}{2007}{417}.

\bibitem{GuntherKruse2010}
\Name{G{\"u}nther S. \and Kruse K.} \REVIEW{Chaos}{20}{2010}{5122}.

\bibitem{Anazawa1992}
\Name{Anazawa T., Yasuda K. \and Ishiwata S.} \REVIEW{Biophys. J.}{61}{1992}{1099}.

\bibitem{Okamura1988}
\Name{Okamura N. \and Ishiwata S.} \REVIEW{J. Muscle Res. Cell M.}{9}{1988}{111}.

\bibitem{Asano2009}
\Name{Asano Y. {\it et al.}} \REVIEW{HFSP J.}{3}{2009}{194}.

\bibitem{JKPJ-PhysReports2007}
\Name{J{\"u}licher F., Kruse K., Prost J. \and Joanny J.-F.} \REVIEW{Phys. Rep.}{449}{2007}{3}.

\bibitem{Kruse2005}
\Name{Kruse K. {\it et al.}}
  \REVIEW{Eur. Phys. J. E}{16}{2005}{5}.

\bibitem{BanerjeeMarchetti2010}
\Name{Banerjee S. \and Marchetti M.~C.} \REVIEW{Soft Matter}{7}{2011}{463}.

\bibitem{Levine2001}
\Name{Levine A.~J. \and Lubensky T.~C.} \REVIEW{Phys. Rev. E}{63}{2001}{041510}.

\bibitem{TanakaLubensky}
\Name{Tanaka T.} \REVIEW{Phys. Rev. Lett.}{40}{1978}{820};
\Name{Golubovi{\'c} L. \and Lubensky T.~C.} \REVIEW{Phys. Rev. Lett.}{63}{1989}{1082}.

\bibitem{StormGardel}
\Name{Storm C. {\it et al.}} \REVIEW{Nature}{435}{2005}{191};
\Name{Gardel M.~L. {\it et al.}} \REVIEW{Science}{304}{2004}{1310};


\bibitem{Parmeggiani2001}
\Name{Parmeggiani A., J{\"u}licher F., Peliti L. \and Prost J.}
  \REVIEW{Europhys. Lett.}{56}{2001}{603}.

\bibitem{Strogatz1994}
\Name{Strogatz S.~H.} \Book{Nonlinear dynamics and chaos} (Westview Press)
  1994.

\bibitem{Magid1985}
\Name{Magid A. \and Law D.} \REVIEW{Science}{230}{1985}{1280}.

\bibitem{Chaudhuri2006}
\Name{O.~Chaudhuri S. H.~P. \and Fletcher D.~A.} \REVIEW{Nature}{445}{2006}{295}.


\bibitem{Thoresen}
\Name{Thoresen T. {\it et al.}} \REVIEW{Biophys. J.}{100}{2011}{446};
\Name{Shimamato Y. {\it et al.}} \REVIEW{Biochem. Biophys. Res. Comm.}{366}{2008}{233};
Dufresne E.~R., Private Comm.

\end{thebibliography}
\end{document}